 \documentstyle[12pt]{article}
 \newcommand{\eq}{\begin{equation}}
 \newcommand{\ee}{\end{equation}}

\begin{document}
 \vspace{1in}
 \begin{center}
 \large\bf
 Network Models of Quantum Percolation \\
 and \\
 Their Field-Theory Representations \\
 \vspace{0.5in}
 \normalsize\rm
 Dung-Hai Lee $^*$ \\
 \vspace{0.2in}
 \em
 IBM Research Division, T. J. Watson Research Center, Yorktown
Heights,
 NY 10598 \\
 \vspace{1.0in}
 \bf
 Abstract
 \end{center}
 \vspace{0.2in}
 \vbox{\narrower\smallskip
 \rm

 We obtain the field-theory representations of several
 network models that are relevant to 2D transport in high
magnetic
 fields. Among them, the simplest one, which is
 relevant to the plateau transition in the quantum Hall effect,
 is equivalent to a particular representation of an
antiferromagnetic
 SU(2N) ($N\rightarrow 0$) spin chain. Since the later can be
mapped
 onto a $\theta\ne 0$, $U(2N)/U(N)\times U(N)$ sigma model, and
since recent
 numerical analyses of the corresponding network give a
delocalization
 transition with $\nu\approx 2.3$, we conclude that the same
exponent is
 applicable to the sigma model.

 \smallskip}
 \vskip 1cm
 \begin{verbatim}
 PACS numbers: 73.50.Jt, 05.30.-d, 74.20.-z
 \end{verbatim}
 \newpage
 \setlength{\baselineskip}{.375in}
 \rm

 In strong magnetic fields, a two-dimensional electron gas
exhibits the
 quantum Hall effect\cite{book} (in this paper we will restrict
ourselves to
 the integer effect) over a {\it wide} range of sample
disorder.\cite
 {jiang}
 The hallmark of the quantum Hall effect is that in the
 neighborhood of a series of `magic' filling factors $f=S_{xy}$,
 $\sigma_{xy}$ exhibits quantized plateaus at
 values $S_{xy} e^2/h$ while $\sigma_{xx}
 \rightarrow 0$.\cite{book}
 Between these filling factors, $\sigma_{xy}$
 interpolates between the plateaus, and
 $\sigma_{xx}$ peaks in the middle of the transitions.
 By analyzing the temperature dependence of the
 widths of the transitions, Wei {\it et al.}\cite{wei}
 have shown convincingly that, if extrapolated to zero
temperature,
 the plateau transition is a continuous phase transition.
 The single diverging length scale at this transition is the
electron
 localization length $\xi\propto\mid B-B_c\mid^{-\nu}$.
 Although Wei {\it et al.}'s results do not give $\nu$ directly,
 a recent experiment using narrow Hall bars suggests
 $\nu\approx 2.3$.\cite{koch} Moreover, considerable numerical
work
 \cite{numerical} has been done for the plateau transition of
 {\it non-interacting} electrons. The result for $\nu$ is
consistent
 with the experimental findings.\cite{evidence}

 It is known that the effective theory for zero-temperature,
 long-wavelength electronic transport in disordered media
 is a non-linear sigma
 model with symplectic symmetry.\cite{efetov}
 In 2D this model has a finite
 localization length (hence gives an insulating state) for
 {\it arbitrarily weak}
 disorders. This result holds when a
 {\it weak}
 magnetic field is applied, where
 the symmetry of the sigma model is
 reduced to {\it unitary}. These results are in apparent
contradiction
 with the quantum Hall effect.
 In trying to reconcile them, Levine, Libby and
Pruisken\cite{llp}
 proposed that
 the proper field theory for electronic transport in {\it strong}
 magnetic fields is a $N\rightarrow 0$, $U(2N)/U(N)\times U(N)$
sigma model
 {\it with a topological term}.
 Subsequently, starting from a microscopic basis,
 Pruisken\cite{pruisken} derived the previously postulated
 sigma model
 \eq
 {\cal L} = {1\over 8}
 \sigma_{xx} Tr[(\partial_{\mu} Q)^2] - {1\over 8} \sigma
 _{xy}\epsilon_{\mu\nu} Tr[Q\partial_{\mu}Q\partial_{\nu}Q],
 \label{action}
 \ee
 where $\mu =x,y$, $\epsilon_{xy}=-\epsilon_{yx}=1$, and
 $\sigma_{\mu\nu}$ are dimensionless conductivities in units of
$e^2/h$.
 In Eq.(1) $Q=u^+\Lambda u$, where $u$ is a 2Nx2N unitary matrix,
and
 $\Lambda$ is a diagonal matrix with $\Lambda_a^a=1$ for $a\leq
N$
 and $\Lambda_a^a=-1$ for $N<a\leq 2N$.
 Pruisken {\it et al.} argued that with the addition of the
second
 (topological) term, Eq.(1) possesses stable
 fixed points at $(\sigma_{xx},\sigma_{xy})=(0,n)$ (corresponding
to
 the quantized plateaus) and critical
 points at $(constant,n+{1\over 2})$ (corresponding
 to the transition points
 between the plateaus).
 Using the sigma model as the paradigm,
 Pruisken conjectured the scaling behavior of
 the plateau transition.\cite{pruisken2} This conjecture was
later verified
 experimentally.\cite{wei}
 Although Pruisken's arguments are compelling, serious
calculations of the
 critical properties are absent.
 It is still an open question as to whether
 Eq.(1) can produce the experimentally or numerically
 observed exponent.\cite{affleck1}

 Quite independently, in an insightful paper, Chalker and
Coddington
 \cite{cc}
 proposed a network model to describe the `quantum percolation'
of the
 semi-classical (equal-potential) orbitals of non-interacting
electrons
 in a strong magnetic field and a
 smooth random potential. Both that work and a recent extension
by
 Lee, Wang and Kivelson\cite{us} show that the network model has
a
 plateau transition with $\nu =2.4\pm 0.2$.
 In brief, the
 network model\cite{cc,us} consists of a checker-board of
low-potential-
 energy plaquettes (marked by `1' in Fig.1a) that are occupied by
 electrons. At the boundary of each
 plaquette, there is an edge state at the
 Fermi energy, in which an electron
 undergoes $\vec{E}\times\vec{B}$ drift in the direction
indicated
 by the arrows. In this model, quantum tunneling takes place
 at the centers of the open squares only. The effect of changing
the
 Fermi energy is reflected in the modification of the tunneling
matrix
 elements. The fact that, in reality, the semi-classical orbitals
have
 different spatial extent is reflected in the {\it random}
Aharonov-Bohm
 phases the electrons accumulate when reaching the tunneling
points.
 In numerical calculations\cite{cc,us}, the localization length
$\xi$ is
 determined in cylindrical systems. Using finite-size scaling the
 thermodynamic $\xi$ was deduced, from which the authors of
Ref.[13]
 concluded that $\nu=2.4\pm 0.2$. Since to calculate $\nu$
 (either numerical or analytically) using Eq.(1) is notoriously
 difficult, one purpose of the present work is to establish a
connection
 between the network model and the sigma model, so that the
exponent
 obtained for the former can be used for the latter.

 Due to a recent experiment on the spin-unresolved plateau
 transition\cite{wei2}, questions concerning the nature of
 plateau transitions in the presence of a strong inter-Landau
 level mixing have been raised. In this paper we generalize
 the network model,\cite {lww} and obtain the corresponding
field-theory
 representation to describe this situation as well.
 Finally, to demonstrate the importance of the random phase on
the
 critical properties of the delocalization transition,
 we consider a network model in its absence. We show that the
 transition in this model belongs to a
 {\it different universality class}.

 A representation of the network model is to view it as a
 one dimensional ($\hat{x}$) stacking of alternating upper-right
 ($\hat{y}$) and lower-left ($-\hat{y}$) moving edge states shown
in
 Fig.1b. Electrons that tunnel at the centers of
 the open squares can be localized
 or extended depending on the assignment of the (random)
tunneling matrix
 elements. To capture the linear dispersion and the chiral nature
of the
 edge states\cite{halperin}, we use the following Hamiltonian
 to model the eigenstates of the network near the Fermi energy
 ($E=0$):
 \eq
 \hat{H}=\sum_x (-1)^{x/a} \int dy {\it v}
 \psi^+ (x,y){{\partial_y}\over i}\psi (x,y)
 -\sum_x\sum_y [ t^\prime (x,y)
 e^{i\phi (x,y)}\psi^+ (x+a,y)\psi (x,y)
 + h.c].
 \ee
 Here $\psi$ is the electron annihilation operator, the $\phi$'s
are the
 random Aharonov-Bohm phases for electrons at the Fermi energy,
 and the $t^\prime$'s are random (positive) tunneling matrix
elements.
 In the second term, the $y$-sum is extended over the
 discrete set of $y$ coordinates at which tunneling takes place.
 Here, without changing the long-distance physics,
 we have assumed that the value of the drift velocity $\it v$
 is uniform. Since the effects of changing $E$ can be absorbed
 into a dependence of $\phi$ and $t^\prime$ on
 $E$, we can concentrate on $E=0$ without losing generality.
 In order to study the critical behavior of the conductivities,
 we consider the following transport action\cite{pruisken}
 \begin{eqnarray}
 {\cal S} & = & \sum_x\int dy i\eta S_p
 \bar{\psi}_p (x,y)\psi_p(x,y) - H \nonumber \\
 & {\rightarrow} & \sum_x\int dy [i\pi\rho\eta S_p\bar{\psi}_p
(x,y)
 \psi_p (x,y)- (-1)^{x/a} \bar{\psi}_p (x,y)
 {{\partial_y}\over i}\psi_p (x,y)] \nonumber \\
 & + & \sum_x\sum_y[
 t(x,y)e^{i\phi (x,y)}\bar{\psi}_p (x+a,y)\psi_p(x,y)
 + h.c.].
 \end{eqnarray}
 In the above, repeated indices $p=\pm$ are summed over,
 $\psi_p$ and $\bar{\psi}_p$
 are Grassmann fields,
 $S_p\equiv sign(p)$, $\rho\equiv 1/(\pi{\it v})$
 is the linear density of states, $\eta$ is a positive
 infinitesimal, and $t\equiv\pi\rho t^\prime$.
 In going from the first to the second lines in
 Eq.(3) we have rescaled the
 fermion fields so as to absorb the edge velocity {\it v}.

 By redefining $\psi_p\rightarrow\psi_p(i\psi_p)$
 and $\bar{\psi}_p\rightarrow
 -i \bar{\psi}_p(\bar{\psi}_p)$ for even (odd) $x/a$'s\cite{note}
 the action becomes
 \begin{eqnarray}
 {\cal S} & = & \sum_x\int dy [(-1)^{x/a} \pi\rho\eta S_p
 \bar{\psi}_p (x,y)\psi_p(x,y)+
 \bar{\psi}_p (x,y)
 \partial_y\psi_p (x,y)] \nonumber \\
 & + & \sum_x\sum_y[
 t(x,y)e^{i\phi (x,y)}\bar{\psi}_p (x+a,y)\psi_p(x,y)
 + h.c.].
 \end{eqnarray}
 Next, we replicate the action and integrate out the random
$\phi$'s and
 $t$'s to obtain
 \begin{eqnarray}
 {\cal S} & = & \sum_x\int dy [(-1)^{x/a} \pi\rho\eta S_p
 \bar{\psi}_{p\alpha}(x,y)\psi_{p\alpha}(x,y)+
 \bar{\psi}_{p\alpha}(x,y)
 \partial_y\psi_{p\alpha}(x,y)] \nonumber \\
 & + & \sum_x\int dy \tilde{F}_x[\bar{\psi}_{p\alpha}(x+a,y)
\psi_{p^\prime\beta}(x+a,y)\bar{\psi}_{p^\prime\beta}(x,y)\psi_{p
\alpha}
 (x,y)].
 \end{eqnarray}
 We note that because $\phi$ is {\it random}, after
 integrating it out only the {\it charge neutral} combinations
 of $\psi$'s and $\psi^+$'s
 are present in Eq.(5). Moreover, the symmetry of the replicated
 action guarantees that only the SU(2N) invariant combinations
appear
 at $\eta=0$.
 As usual, the repeated replica indices $\alpha,\beta =1...N$ are
summed
 over, and $\tilde{F}_x (Z)={1\over {a}}<t^2(x)> Z + ...$.
 We note that $\tilde{F}$
 has an explicit $x$-dependence. It reflects
 the fact that the electrons tunnel across the
occupied/unoccupied
 regions
 for the even-odd/odd-even columns. As a result, the
corresponding
 $<t^2>$ are different.

 If we view $y$ as the imaginary time $\tau$,
 Eq.(5) is the coherent-state
 path-integral action of a 1D quantum field theory
 described
 by the following Hamiltonian
 \eq
 \hat{H}=\sum_x [(-1)^{x/a} \pi\rho\eta S_p
 {\psi^+_{p\alpha}}(x)\psi_{p\alpha}(x)]
 +\sum_x F_x[\psi^+_{p\alpha} (x+a)
 \psi_{p^\prime\beta}
(x+a)\psi^+_{p^\prime\beta}(x)\psi_{p\alpha}
 (x)].
 \ee
 The relation between $F$ in Eq.(6) and $\tilde{F}$ in Eq.(5) is
that
 upon normal ordering $F\rightarrow \tilde{F}$.
 The relation to the SU(2N) spin chain becomes explicit after we
 rewrite Eq.(6) in terms of the SU(2N) generators
 $\hat{S}_a^b\equiv\psi_a^+\psi_b-\delta_{ab}
 {1\over {2N}}\sum_c\psi_c^+\psi_c$,
 where $a\equiv (p,\alpha)$ and takes on $2N$ values.
 The final spin Hamiltonian is
 \eq
 \hat{H}=\sum_x (-1)^{x/a} \pi\rho\eta Tr[\Lambda
\hat{S}(x)]+\sum_x F_x(
 Tr[\hat{S}(x+a)\hat{S}(x)]),
 \ee
 where $\Lambda$ is the diagonal c-number matrix defined earlier,
 and $Tr[\Lambda \hat{S}]\equiv\sum_{a,b}\Lambda_a^b\hat{S}_b^a$,
 $Tr[\hat{S}\hat{S}^\prime]\equiv\sum_{a,b}
 \hat{S}_a^b\hat{S^\prime}_b^a$.
 Since $\hat{H}$ commutes with the site occupation number
$n(x)=\sum_a
 \psi^+_a(x)\psi_a(x)$, Hilbert spaces corresponding to different
 $\{n(x)\}$ decouple. The ground state of Eq.(7) lies in the
Hilbert
 space where $n(x)=N$ for all x. In this Hilbert space,
 a particular representation for SU(2N) is realized. This
representation
 is characterized by a Young tableau with a single column of
length N.

 Analogous to the SU(2) quantum spin chain, which in the
large-spin
 limit is equivalent to the U(2)/U(1)xU(1)=O(3) sigma model,
 one can show that at $\eta=0$ the SU(2N) spin chain of Eq.(7)
 is equivalent to the following $U(2N)/U(N)\times U(N)$ sigma
model
 in the large representation limit\cite{large,affleck2}
 \eq
 {\cal L} =
 {M\over {16}}(\sqrt{1-R^2}) Tr (\partial_{\mu}Q)^2
 +{M\over {16}}(1-R)\epsilon
 _{\mu\nu}Tr[Q\partial_{\mu}Q\partial_{\nu}Q],
 \ee
 where $\mu =\tau,x$, and
 $R\equiv [F^\prime_{x+a}(Z_0)-F^\prime_x (Z_0)]/[F^\prime_{x+a}
 (Z_0)+F^\prime_x (Z_0)]$ (here $Z_0\equiv -NM^2/4$ and
 $F^\prime (Z)\equiv dF/dZ$).
 For the network model $M=1$, and Eq.(8) is
 massless at $R=0$. For $R\ne 0$ the system remains massive.
 Thus we identify the transition from $R<0$ to $R>0$ as the
plateau
 transition.
 At the critical point ($R=0$) the $\hat{x}$-translational
 symmetry is restored. If we {\it assume} that a) the mapping
from Eq.(7)
 to Eq.(8) is valid down to
 $M=1$, and b) at
 the critical point the system is described by a
 translationally-invariant SU(2N) spin chain with M=1, then
 by comparing Eq.(1) and Eq.(8) we conclude that
 $\sigma_{xx}=\sigma_{xy}=1/2$ at the critical
point.\cite{us,klz}
 In any case, {\it if} the density of states stays finite at the
 transition,\cite{dos} then
 the dimension of $Q$ remains zero.
 As a result, the
 $\sigma_{\mu\nu}$ in Eq.(1) are dimensionless, and hence are
generically
 universal at the transition.\cite{klz}

 Modifications can be made so that the network model represents
 other physically relevant situations. For example,
 to study the effects of neighboring
 Landau-level mixing on the plateau transition
 we introduce {\it two} edge states
 with the {\it same}
 circulation in each plaquette of Fig.1.\cite{lww}
 The Hamiltonian including the inter-`channel' mixing is given by
 \begin{eqnarray}
 \hat{H} & = & \sum_x (-1)^{x/a} \int dy {\it v}_{\sigma}
 \psi^+_{\sigma} (x,y){{\partial_y}\over i}\psi_{\sigma} (x,y)
 -\sum_x\sum_y [ t^\prime_{\sigma} (x,y)
 e^{i\phi_{\sigma}(x,y)}\psi^+_{\sigma} (x+a,y)\psi_{\sigma}
(x,y)
 + h.c] \nonumber \\
 & - & \sum_x\sum_y [ t^\prime_{12} (x,y)
 e^{i\phi_{12}(x,y)}\psi^+_1 (x+a,y)\psi_2 (x,y)
 + h.c].
 \end{eqnarray}
 Here $\sigma =1,2$ is the channel index and is implicitly summed
over.
 Straightforward generalization of the steps between Eq.(2)
 and Eq.(7) now gives
 \eq
 \hat{H}=\sum_x (-1)^{x/a} \pi\rho_{\sigma}
 \eta Tr[\Lambda \hat{S}_{\sigma}(x)]+\sum_x F_{x,\sigma}
 (Tr[\hat{S}_{\sigma}(x+a)\hat{S}_{\sigma}(x)])
 +\sum_x F_{x,12}(Tr[\hat{S}_1(x)\hat{S}_2(x)])
 \ee
 In the above $F_{x,\sigma}(Z)=
 {{\pi^2\rho_{\sigma}^2}\over {a}}<{t^\prime}_{\sigma}^2(x)> Z
 + ...$, and are {\it antiferromagnetic} interactions. On the
contrary,
 $F_{x,12}(Z)=-
 {{\pi^2\rho_1\rho_2}\over {a}}<{t^\prime}_{12}^2(x)> Z + ...$,
and is
 {\it ferromagnetic}.
 Therefore inter-Landau mixing causes a ferromagnetic coupling
 between two otherwise decoupled
 SU(2N) spin chains.

 In the absence of the inter-chain coupling there are two
 transitions as we tune $R_1$ and $R_2$ keeping $R_1>R_2$.
 ($R_{\sigma}\equiv
[F^\prime_{x+a,\sigma}(Z_0)-F^\prime_{x,\sigma}
 (Z_0)]/[ F^\prime_{x+a,\sigma}
 (Z_0)+F^\prime_{x,\sigma} (Z_0)]$)
 All three massive phases break the translational symmetry and
 correspond to the (even-odd, even-odd), (even-odd, odd-even),
and
 (odd-even, odd-even) `spin-Peierls' phases.
 In the presence of a {\it strong}
 ferromagnetic inter-chain coupling we expect
 vertical spin pairs to form the $M=2$ representation of SU(2N).
At long
 wave-length, the problem is equivalent to a {\it single}
 antiferromagnetic spin chain in the $M=2$ representation. In
this case
 there are also two transitions as we tune $R_1$ and $R_2$. Among
the
 three massive phases, two break translation symmetry and they
 correspond to the (even-odd) and (odd-even) spin-Peierls phases.
 The phase which remains translationally invariant is the SU(2N)
 analog of the Haldane phase.\cite{aklt}
 Thus the phase structure and the universality class of the phase
 transition are {\it preserved} in the presence of a strong
 inter-Landau level mixing. Although we have not done
 calculations for intermediate inter-chain couplings, based on
 the knowledge about the SU(2) spin chains\cite{solyom} we
 expect the same behavior as in the zero and strong coupling
limits.

 In addition to the quantum spin chains, the network model
contains
 another interesting field theory.
 In specific, if we set the random
 phases $\theta =0$, hold $t(x+a,y)+t(x,y)$ fixed and let
 $t(x+a,y)-t(x,y)$ be random, the network model is equivalent to
 a 2N-component Gross-Neveu model\cite{gn} in the limit
 $N\rightarrow 0$.
 Since the latter model has a different localization
 length exponent, we conclude that the random phase is essential
 in determining the universality class of the delocalization
transition.
 To obtain the field theory we
 group each adjacent upper-right and
 lower-left moving edge states into a doublet. For each doublet
 we define $\psi_+$ and $\psi_-$ as the annihilation operators
for
 the upper-right and lower-left moving electrons respectively.
 Moreover, we construct a Dirac spinor $\Psi$ such that
 $\Psi=\pmatrix{\psi_+\cr \psi_-\cr}$.
 It is simple to show that in the continuum limit Eq.(2) reduces
to
 \eq
 H=\sum_x\int dy \Psi^+(x,y)\{\sigma_z p_y
 +[(t(x+1,y)-t(x,y)]\sigma_x
 +a[t(x+a,y)+t(x,y)]\sigma_y p_x\}\Psi(x,y).
 \ee
 Here $p_{\mu}={{\partial_{\mu}}\over i}$ ($\mu =x,y$),
 and $\sigma_{\mu}$ are the Pauli matrices.
 In obtaining the above
 we have flipped the signs of $\Psi$ and $\Psi^+$ for every
 other doublets.
 After going to the spinor basis that diagonalize $\sigma_y$, we
let
 $\sum_x\rightarrow\int{{dx}\over {2a}}$,
 ${{t(x,y)+t(x+a,y)}\over 2}y\rightarrow y$ and $x/2a\rightarrow
x$.
 As the result, we
 obtain
 \eq
 H=\int dxdy
\Psi^+(x,y)[\sigma_{\mu}p_{\mu}+m(x,y)\sigma_z]\Psi(x,y),
 \ee
 where $m(x,y)=2[t(x+a,y)-t(x,y)]/[t(x+a,y)+t(x,y)]$.
 We recognize that Eq.(12) is the Hamiltonian for 2D Dirac
electrons
 with random masses. To study the conductivities we study the
 following $E=0$ transport action
 \eq
 {\cal S}=\int d^2x
\bar{\Psi}_p[\gamma_{\mu}\partial_{\mu}+m-i\eta
 S_p\sigma_z ] \Psi_p.
 \ee
 Here we have let $x_{\mu}\rightarrow\epsilon_{\mu\nu}x_{\nu}$,
and
 redefined $\bar{\Psi}$ so that $\bar{\Psi}\rightarrow
-\bar{\Psi}
 \sigma_z$.
 Finally, we replicate and integrating out $\delta m =m-<m>$ to
obtain:
 \eq
 {\cal S}=\int d^2x \bar{\Psi}_{p\alpha}
 [\gamma_{\mu}\partial_{\mu}+<m>-i\eta
 S_p\sigma_z]\Psi_{p\alpha}-{g\over 2}
 (\bar{\Psi}_{p\alpha}\Psi_{p\alpha})
 (\bar{\Psi}_{p^\prime\beta}\Psi_{p^\prime\beta}),
 \ee
 where $g=<(\delta m)^2>$. In the limit $\eta\rightarrow 0$,
Eq.(14)
 describes the 2N-component Gross-Neveu model.
 We have checked numerically\cite{lw} that for small $<\delta
m^2>$
 the phase transition of the network model
 has $\nu =1$, consistent with the result of the perturbative
 analysis of Eq.(14).\cite{shankar}

 Acknowledgments: We thank E. Fradkin, S. Kivelson, N. Read, Z.
Wang
 and X-G Wen for useful discussions. We also thank J. Chalker for
 communicating his numerical results to us prior to their
publication.
 Part of the work was initiated at the Aspen Center for Physics.

 \noindent[
 * New address: Department of physics, University of California at Berkeley,
 Berkeley, CA 94720]

 \newpage
  \bibliographystyle{unsrt}

\begin{thebibliography}{99}
  \bibitem{book}
  See e.g. ``The Quantum Hall Effect'', 2nd edition, edited
  by R.E. Prange and S.M. Girvin, Springer-Verlag (1990).
  \bibitem{jiang}
  H.W. Jiang, C.E. Johnson, K.L. Wang, and S.T. Hannahs,
  Phys. Rev. Lett. {\bf71}, 1439 (1993).
  \bibitem{wei}
  H.P. Wei, D.C. Tsui, M. Paalanen, and A.M.M. Pruisken,
  Phys. Rev. Lett. {\bf61}, 1294 (1988).
  \bibitem{koch}
  S. Koch, R. Haug, K. v. Klitzing, and K. Ploog,
  Phys. Rev. Lett. {\bf67}, 883 (1991).
  \bibitem{numerical}
  B. Huckestein and B. Kramer, Phys. Rev. Lett. {\bf64}, 1437
(1990);
  B. Mieck, Europhysics Lett. {\bf13}, 453 (1990); Y. Huo and
R.N. Bhatt,
  Phys. Rev. Lett. {\bf68}, 1375 (1992); D. Liu and S. Das Sarma,
  Mod. Phys. Lett. B {\bf7}, 449 (1993).
  \bibitem{evidence}
  One can take this agreement as an experimental indication that
the
  Coulomb interaction is {\it irrelevant} at the plateau
transition.
  Nonetheless,
  the rigorous theoretical justification for such irrelevance
  is still lacking.
  \bibitem{efetov}
  F.Wegner, Z. Physik B {\bf36},209 (1979);K. Efetov, A. Larkin
and
  D. Khemelnitskii, JETP {\bf52}, 568 (1980).
  \bibitem{llp}
  H. Levine, S.B. Libby and A.M.M. Pruisken, Phys. Rev. Lett.
{\bf51},
  1915 (1983).
  \bibitem{pruisken}
  A.M.M. Pruisken, Nucl. Phys. B {\bf235} FS[11], 277 (1984), and
  chapter 5 of Ref.[1].
  \bibitem{pruisken2}
  A.M.M. Pruisken, Phys. Rev. Lett. {\bf61}, 1297 (1988).
  \bibitem{affleck1}
  I. Affleck, Nucl. Phys. B {\bf265}, 409 (1986).
  \bibitem{cc}
  J.T. Chalker and P.D. Coddington, J. Phys. C {\bf21}, 2665
(1988).
  \bibitem{us}
  D-H Lee, Z. Wang and S.A. Kivelson, Phys. Rev. Lett. {\bf70},
4130
  (1993).
  \bibitem{wei2}
  H.P Wei, S.W. Hwang, D.C. Tsui, and A.M.M. Pruisken
  Surf. Sci., {\bf229} 34 (1990).
  \bibitem{lww}
  For the numerical results on these new networks see
  D-H Lee, Z. Wang, and X-G Wen, to be published; J. Chalker
  to be published.
  \bibitem{halperin}
  B.I. Halperin, Phys. Rev. B {\bf25}, 2185 (1982).
  \bibitem{note}
  Since $\psi$ and $\bar{\psi}$ are {\it independent} Grassmann
  fields, they can be transformed independently.
  \bibitem{large}
  The large representations are characterized by Young tableaus
with
  $M$ columns ($M>>1$) of length $N$.
  \bibitem{affleck2}
  I. Affleck, Nucl. Phys. B {\bf257} FS[14], 409 (1986); D-H Lee
to
  be published.
  \bibitem{dos} In the transfer matrix calculation, one only computes
  the connected part of the spin-spin correlation function. Hence it
  bares no implication on whether the spin operator itself has nonzero
  expectaion value (or the density of state is finite) at the transition.
  This question is currently under
  investigation. For relevant works on this issue see, e.g. E. Fradkin,
  Phys. Rev. B {\bf33}, 3257 (1986); and ibid 3263 (1986).
  \bibitem{klz}
  S.A. Kivelson, D-H Lee and S-C Zhang, Phys. Rev. B {\bf46},
2223
  (1992).
  Y.Huo, R.E. Hetzel and R.N. Bhatt, Phys. Rev. Lett. {\bf70},
481
  (1993).
  \bibitem{aklt}
  F.D.M. Haldane, Phys. Lett. {\bf93}A, 464 (1983); Phys. Rev.
Lett. {\bf
  50}, 1153 (1983); J. Appl. Phys. {\bf57}, 3359 (1985); I.
Affleck,
  Nucl. Phys. B{\bf257}, 397 (1985).
  \bibitem{solyom} J. Solyom and J. Timonen, Phys. Rev. B
{\bf34},
  487 (1986).
  \bibitem{gn}
  D.J. Gross and A. Neveu, Phys. Rev. D {\bf10}, 3235 (1974)
  \bibitem{lw}
  Z. Wang and D-H Lee, to be published.
  \bibitem{shankar}
  See e.g. R. Shankar, BCSPIN lectures, Kathmandu, 1991.

  \newpage
  \centerline{FIGURE CAPTIONS}

  \ \

  \noindent{\bf Figure 1}. a) The network model.
  Here the arrows indicate the direction of
  the edge velocity, and the open squares enclose the tunneling
  points.
  b) A representation of the network model as
  alternating $\hat{y}$ and $-\hat{y}$-moving fermions
  that tunnel at the centers of the squares.

   \ \

 \vspace*{\fill}
 \end{thebibliography}
  
 \end{document}